\documentclass[sn-mathphys]{sn-jnl}

\jyear{2021}%

\theoremstyle{thmstyleone}%
\theoremstyle{thmstyletwo}%

\theoremstyle{thmstylethree}%

\usepackage{indentfirst}
\usepackage[switch]{lineno}
\begin{document}

\title{Design and simulation of a novel 4H-SiC LGAD timing device}


\author[1,2]{\fnm{Keqi} \sur{Wang}}\email{wangkq@ihep.ac.cn}
\equalcont{These authors contributed equally to this work.}

\author[1,3]{\fnm{Tao} \sur{Yang}}\email{yangtao@ihep.ac.cn}
\equalcont{These authors contributed equally to this work.}

\author[1,3,4]{\fnm{Chenxi} \sur{Fu}}

\author[2]{\fnm{Li} \sur{Gong}}

\author[5]{\fnm{Songting} \sur{Jiang}}

\author[2]{\fnm{Xiaoshen} \sur{Kang}}

\author[1,3,6]{\fnm{Zaiyi} \sur{Li}}

\author[4]{\fnm{Hangrui} \sur{Shi}}

\author[1]{\fnm{Xin} \sur{Shi}}

\author[4]{\fnm{Weimin} \sur{Song}}

\author[1]{\fnm{Congcong} \sur{Wang}}

\author[7,8]{\fnm{Suyu} \sur{Xiao}}

\author[1]{\fnm{Zijun} \sur{Xu}}

\author*[1]{\fnm{Xiyuan} \sur{Zhang}}\email{zhangxiyuan@ihep.ac.cn}

\affil*[1]{\orgdiv{Institution of High Energy Physics}, \orgname{Chinese Academy of Sciences}, \orgaddress{\city{Beijing}, \postcode{100049},  \country{China}}}

\affil[2]{\orgdiv{Department of Physics}, \orgname{Liaoning University}, \orgaddress{\city{Shenyang}, \postcode{110036}, \state{Liaoning}, \country{China}}}

\affil[3]{\orgname{University of Chinese Academy of Sciences}, \orgaddress{\city{Beijing}, \postcode{100049},  \country{China}}}

\affil[4]{\orgdiv{Department of Physics},\orgname{Jilin University}, \orgaddress{\city{Changchun}, \postcode{130015}, \state{Jilin}, \country{China}}}

\affil[5]{\orgdiv{Department of Mechanical Engineering},\orgname{Heilongjiang University Of Science And Technology}, \orgaddress{\city{Harbin}, \postcode{150022}, \state{Heilongjiang}, \country{China}}}

\affil[6]{\orgdiv{College of Physics and Electronic Engineering}, \orgname{Shanxi University}, \orgaddress{\city{Taiyuan}, \postcode{030006}, \state{Shanxi}, \country{China}}}

\affil[7]{\orgname{Shandong Institute of Advanced Technology}, \orgaddress{\city{Jinan}, \postcode{250100},  \country{China}}}

\affil[8]{\orgname{Shandong University}, \orgaddress{\city{Jinan}, \postcode{250100},  \country{China}}}

\abstract
{Silicon-based fast time detectors have been widely used in high energy physics, nuclear physics, space exploration and other fields in recent years. However, silicon detectors often require complex low-temperature systems when operating in irradiation environment, and their detection performance decrease with the increase of irradiation dose. Compared with silicon, silicon carbide (SiC) has a wider bandgap, higher atomic displacement energy, saturated electron drift velocity and thermal conductivity. Simultaneously, the low gain avalanche detector avoids crosstalk and high noise from high multiplication due to its moderate gain, and thus can maintain a high detector signal without increasing noise. Thus, the 4H-SiC particle detector, especially the low gain avalanche detector has the potential to detect the minimal ionized particles (MIPs) under extreme irradiation and high temperature environments.\par
In this work, the emphasis was placed on the design of a 4H-SiC Low Gain Avalanche Detector (LGAD), especially the epitaxial structure and technical process which played the main roles. In addition, a simulation tool--RASER(RAdiation SEmiconductoR) was developed to simulate the performances including the electrical properties and time resolution of the 4H-SiC LGAD we proposed. The working voltage and gain effectiveness of the LGAD were verified by the simulation of electrical performances. The time resolution of the LGAD is (35.0 ± 0.2) ps under the electrical field of -800 V, which is better than that of the 4H-SiC PIN detector.}


\maketitle

\section{Introduction}\label{sec1}

Low gain avalanche detector (LGAD), as a novel detector technology, has attracted extensive attention in recent years. By adding a gain layer under the cathode of the PIN detector, a collision-ionization multiplication of the electron-hole pair occurs in the gain layer, resulting in excellent timing resolution\cite{ref-LGAD&PIN}. In particular, for the requirements of the High Granularity Timing Detector (HGTD) in ATLAS and the End-cap Timing Layer (ETL) in CMS experiments, silicon-based LGAD with 50 ps time resolution has been developed which can resist the radiation of 2×$10^{15}~n_{eq}$/cm$^{2}$\cite{ref-Si_LGAD1,ref-Si_LGAD2,ref-Si_LGAD3,ref-Si_LGAD4}. However, it has been reported\cite{ref-irradiation} that the charge collection efficiency (CCE) and time resolution of silicon-based LGAD will deteriorate with increasing irradiation flux and will not work normally under irradiation above 2×$10^{15}~n_{eq}$/cm$^{2}$. In addition, in order to reduce power and improve the signal-to-noise ratio (SNR) of semiconductor detectors, silicon-based LGAD devices need to operate at a low temperature of -30 ℃\cite{ref-low temperature1, ref-low temperature2}, which greatly increases the operation cost of detectors. Moreover, the low temperature cooling system not only increases volume of the detector system, but also increases the difficulty of maintaining the stability of the detector system. Therefore, there is a high demand for research into a semiconductor detector with strong radiation resistance, excellent time resolution and stable operation at room temperature or even high temperature.\par
Silicon carbide (SiC), as the third generation semiconductor, has been investigated as a detector material for nearly 20 years\cite{ref- SiC material}. In recent years, SiC has received more attention due to the interest of semiconductor industries in the renewable energy revolution, leading to  broader application in power-efficient transistors, photovoltaic inverters and electric vehicle drive trains, etc. Compared with silicon, SiC has higher band-gap and atomic displacement, which leads to its strong potential for irradiation resistance\cite{ref-radiation}. In addition, the higher breakdown electric field, saturated electron drift velocity and thermal conductivity mean that a SiC detector would have a faster time response and lower temperature sensitivity\cite{ref-temperature}.  Zhang's group reported a 4H-SiC detector with a time response of 117 ps to alpha particles\cite{ref-Zhang time response}. However, there are very few reports on the time response of Minimum Ionizing Particles (MIPs). We have investigated the time performance of a 4H-SiC PIN detector with a thickness of 100$ ~\mu m$, and the time resolution of MIPs is 94 ps\cite{ref-sic_pin}. It is noteworthy that the average ionization rates of electron-hole pairs of MIPs in 4H-SiC and Si are 55 pairs/$\mu m$ and 75 pairs/$\mu m$, respectively, which means that the response of the 4H-SiC detector is less than that of the Si detector. At present, the main problem of Si-based LGAD is the reduction of timing performance caused by high-intensity irradiation. Once 4H-SiC has the excellent timing performance of LGAD, combined with its irradiation resistance and temperature stability, 4H-SiC LGAD will be a candidate for high pileup, high radiation environment and stable operation at room temperature. \par
In this work, a novel 4H-SiC LGAD timing device was designed by adjusting the doping concentration and epitaxial structure using the RAdiation SEmiconductoR (RASER) software, and its electrical characteristics were simulated. The design of the technical process has also been carried out and RASER is further used to simulate the time resolution of the 4H-SiC LGAD. The time resolution of the 4H-SiC LGAD designed 
by adjusting the thickness and doping concentration is basically the same as that of the Si-based LGAD detector (34 ps)\cite{ref-si_lgad}, which is obviously superior to the corresponding PIN device\cite{ref-sic_pin}.

\section{Design scheme of 4H-SiC LGAD}\label{sec2}

In this work, the design scheme of 4H-SiC LGAD will be introduced from two aspects: epitaxial structure and technical process.

\subsection{Epitaxial structure of 4H-SiC LGAD}\label{subsec2}

\indent In this work, it is proposed to design a 4H-SiC LGAD timing device by full epitaxial growth. Different from Si LGAD, the hole multiplication rate in 4H-SiC is greater than that of electrons. To achieve more excellent carrier multiplication, the N-type epitaxial layers including bulk layer (N$^-$) and gain layer(N$^+$) has been selected. In detail, due to the high average ionization energy of the 4H-SiC, a relatively thicker bulk layer should be required to generate more initial ionized electron-hole pairs for incident particles, which can be read out by electronic readout system after undergoing 10 times of low-gain multiplication (\textgreater 4fC). Therefore, the thickness of the bulk layer of the 4H-SiC LGAD should be at least 50$ ~\mu m$. In addition, the doping concentration of the bulk layer with the thickness of 50$ ~\mu m$ should be as low as possible in order to deplete the whole bulk region at a certain operating voltage. At present, the epitaxial growth of 4H-SiC with a doping concentration of 10$^{14}$~cm$^{-3}$ can be realized in industry.\par
To achieve low-gain multiplication of 4H-SiC LGAD devices, an N$^+$ gain layer should be epitaxially grown above the bulk layer. Under the reverse bias, a sharply high electric field can be generated in this region, causing the electron-hole pairs generated by particles passing through the device to be collision-ionized multiplied in this thin gain layer. Based on the experience of Si LGAD, a thickness of the gain layer is from 0.1$ ~\mu m$ to 1$ ~\mu m$. In this work, 1$ ~\mu m$ thickness gain layer with better uniformity has been selected. Meanwhile, the doping concentration range is 1.4×10$^{17}$~cm$^{-3}$ $\sim$ 1.48×10$^{17}$~cm$^{-3}$, which restrict the corresponding multiplication coefficient and meet the industrial epitaxial process requirements. In addition, the P$^{++}$ layer with a doping concentration of 5×10$^{19}$~cm$^{-3}$ and a thickness of 0.3$ ~\mu m$ should be epitaxially grown on the N$^+$ gain layer. The highly doped P$^{++}$ layer is easy to form ohmic contact, improving its charge collection efficiency and time resolution. \par
Overall, the epitaxial structure of 4H-SiC LGAD is $P^{++}/N^+_{gain}/N^-_{bulk} /N_{buffer}/N^{++}_{substrate}$ as shown in Figure 1(a). Besides, the aluminum ions and nitrogen ions were used as p-doping and n-doping, respectively. The epitaxial structure of 4H-SiC LGAD was prepared based on the above design. In order to prove the technological feasibility of the design scheme and the accuracy of epitaxial parameters such as doping concentration and thickness, secondary ion mass spectrometry (SIMS) has been carried out for characterization at a depth of 3 $\mu m$, and the results are shown as Figure 1(b). The actual thickness and doping concentration of the device were consistent with the corresponding designed parameters, which provided support for our subsequent research on the time resolution of 4H-SiC LGAD. 

\begin{figure}[H]
\centering
\includegraphics[trim=0 40 0 20 ,scale=0.48]{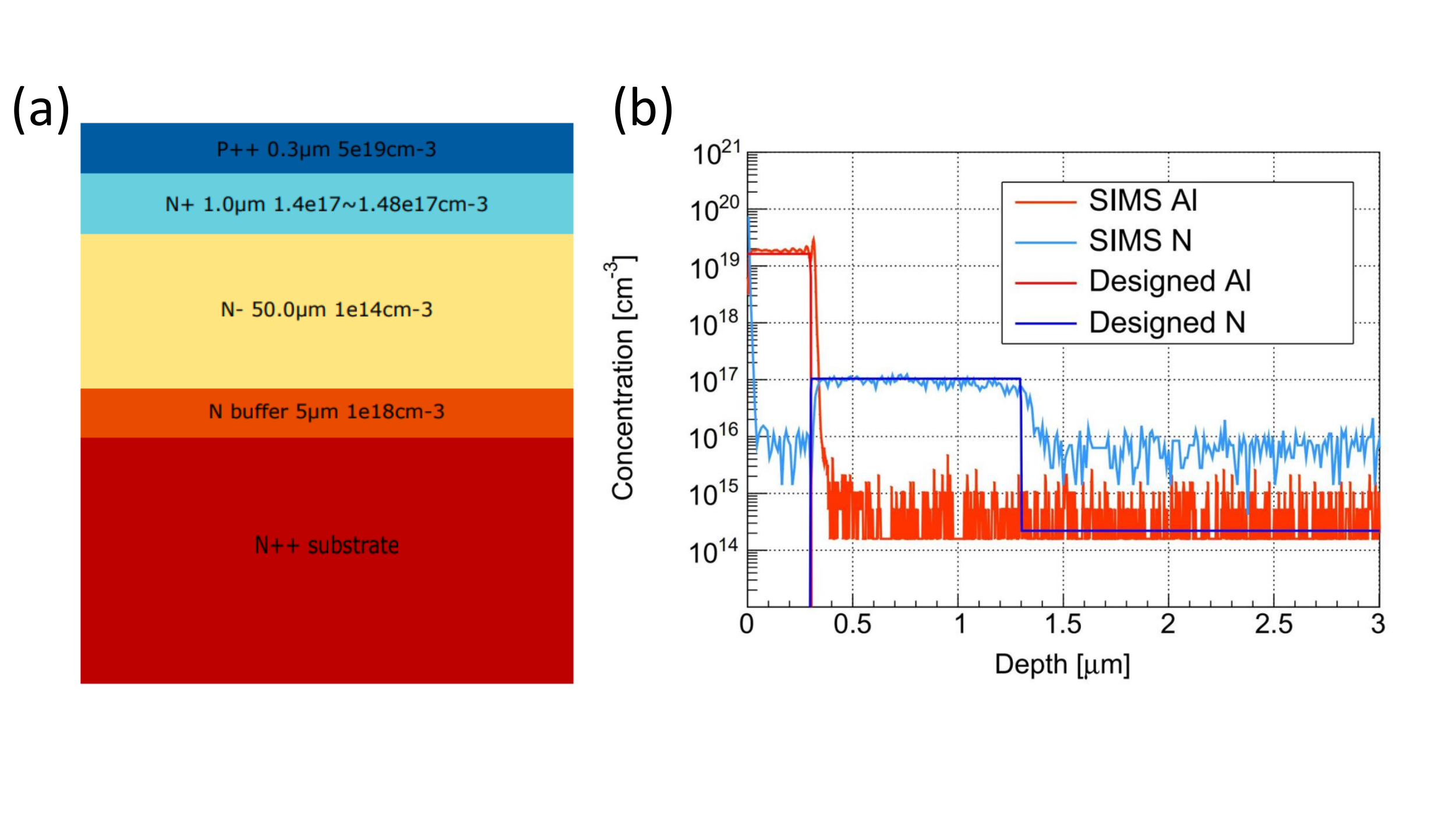}
\caption{(a) Schematic diagram of SiC LGAD epitaxial structure with doping concentration and thickness. (b) Comparison of SIMS measurement of doping concentration and thickness and the designed parameters.\label{fig1}}
\end{figure}   
\unskip

\subsection{Technical processes}\label{subsec2}

It has been reported that the breakdown voltage of conventional planar 4H-SiC devices is mainly determined by the local electric field caused by the device boundary effect \cite{ref-breakdown voltage}. For the 4H-SiC LGAD designed in this work, the high electric field region generated in the gain layer is close to the device  surface, and the boundary effect is particularly obvious, resulting in more serious premature breakdown. In addition, the operating voltage of the 4H-SiC LGAD needs to be greater than the full depletion voltage to work properly, which means that the operating voltage should be relatively high. Therefore, ensuring that the device does not break down in advance is a major consideration for the technical process. Etching the terminal junction is an effective method to avoid premature breakdown caused by the boundary effect of high electric field\cite{ref-etch}. In this work, etching table with 7° has been used and the technical process designed on the existing epitaxial wafer is shown in Figure 2(a). In addition, the design and preparation of P-type ohmic contact electrodes are also very important to improve the performance of 4H-SiC LGAD. High temperature anneal alloy system based on Ni/Ti/Al has been widely used to produce P-type ohmic contact\cite{ref-ohmic1,ref-ohmic2}, with specific contact resistance up to 1.8×10$^{-5}~{\Omega}~cm^2$. The same alloy system is used to make the P-type ohmic contact electrodes and N-type ohmic contact electrodes and the annealing temperature is 800 °C\cite{ref-anneal}. A low resistivity ohmic contact can be formed between the metal and silicon carbide. The anode is N$^{++}$ layer, and the cathode is P$^{++}$ layer. Finally, in order to prevent gas from oxidizing the wafer, a layer of silica has been deposited as a passivation on the exposed part of electrodes. By comparing the thickness of silica and the laser wavelength, the silica with a thickness of 364 nm is finally determined. It not only provides protection, but also uses as a transparency-enhancing film for subsequent laser tests. The first version of the 4H-SiC LGAD device is named SICAR(Silicon CARbide) by our group.\par
To implement the above process, three masks has been designed, including 8 structures (SICAR1-1$\sim$8), which are shown in Figure 2(b). Mask 1 is used to determine the etching-table structure, Mask 2 is used to determine the p-type electrodes and Mask 3 is used to determine the electrode pads of the 4H-SiC LGAD. To investigate the effects of size and shape on the electrical performance and time resolution, these 8 devices with different sizes and corner radii has been designed. Specifically, the shape and electrodes of SICAR1-1 (5000 × 5000$ ~\mu m$) refer to the existing 4H-SiC PIN detector\cite{ref-sic_pin}. SICAR1-2 (1000 × 1000$ ~\mu m$), a 5x5 array of SICAR1-6 is designed to measure the position resolution. Moreover, in order to study the multiplication heterogeneity of the 4H-SiC LGAD, a laser incidence hole with radius of (25$ ~\mu m$) is reserved on SICAR1-3 (1000 × 1000$ ~\mu m$). Devices of SICAR1-4 (1000 × 1000$ ~\mu m$) and SICAR1-5 (1000 × 1000$ ~\mu m$) are 2×2 arrays with array spacing of 50$~\mu m$ and 100$~\mu m$, respectively, which are designed to study the size of detector "dead zone" under different array spacing. In order to study the influence of corner radius on leakage current, 100$ ~\mu m$, 200$ ~\mu m$ and 500$ ~\mu m$ corner radius are designed for SICAR1-6 (1000 × 1000$ ~\mu m$), SICAR1-7 (1000 × 1000 $ ~\mu m$) and SICAR1-8 (1000 × 1000$ ~\mu m$) devices, respectively.

\begin{table}[h]
\begin{center}
\begin{minipage}{174pt}
\caption{The layout mask contains the device structure.}\label{tab1}%
\begin{tabular}{@{}llll@{}}
\toprule
Label & Type  & Radius~$[\mu m]$\\
\midrule
SICAR1-1      &Single	& 500  \\
SICAR1-2      &$5\times5$  & 100  \\
SICAR1-3      &Single	& 100  \\
SICAR1-4      &$2\times2$	& 100  \\
SICAR1-5      &$2\times2$	& 100  \\
SICAR1-6      &Single	& 100  \\
SICAR1-7      &Single	& 200  \\
SICAR1-8      &Single	& 500  \\
\botrule
\end{tabular}
\end{minipage}
\end{center}
\end{table}

\begin{figure}[H]
\centering
\includegraphics[trim=0 40 0 20 ,scale=0.48]{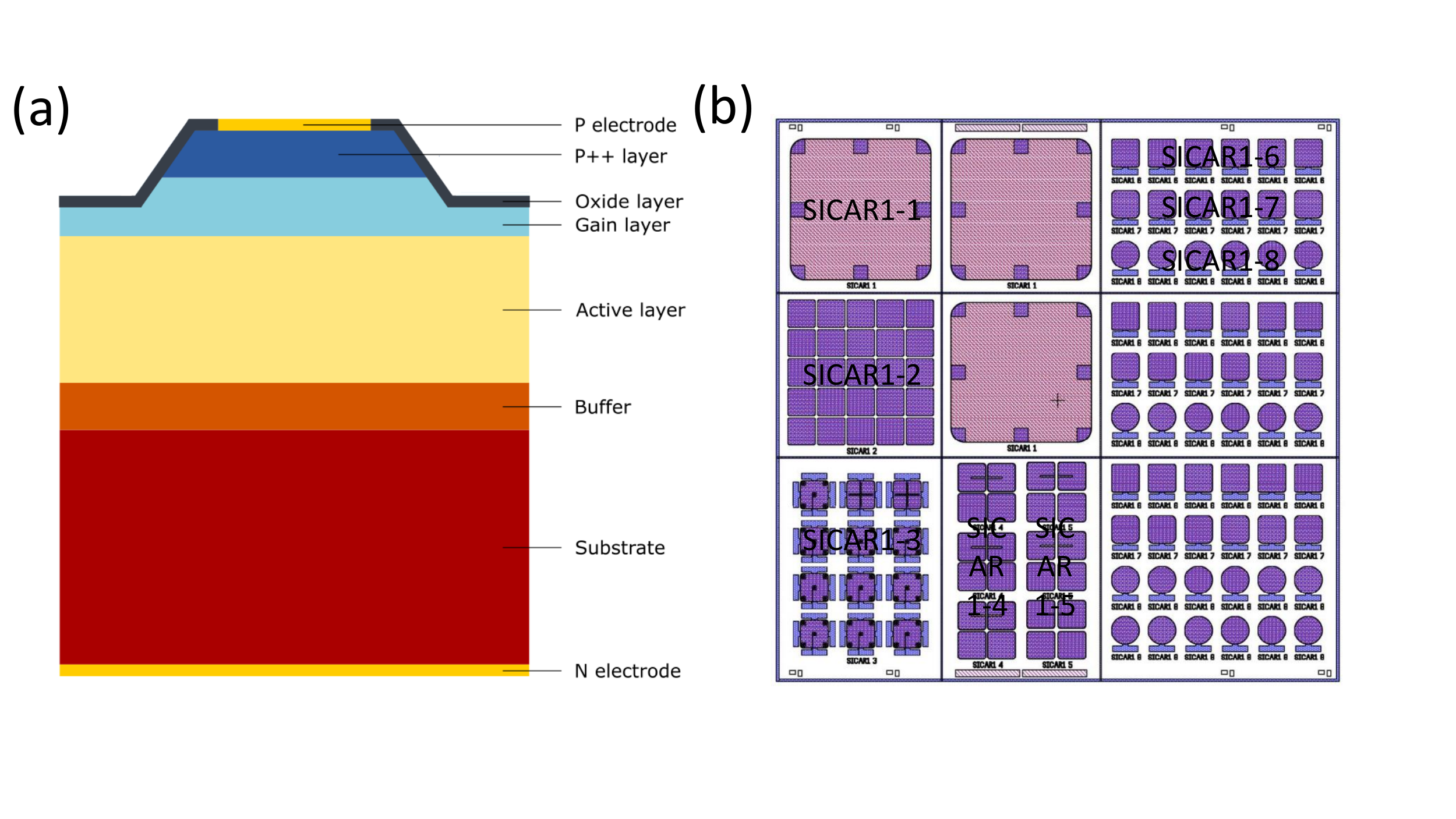}
\caption{(a) Sketch of the structure of an LGAD (not to scale).(b) SiC LGAD 2×2 $cm^2$ layout mask. Different size and different rounded corner radius pads were allocated and aligned in this mask.\label{fig2}}
\end{figure}   
\unskip

\section{RASER}\label{sec3}
To estimate the I-V and C-V characteristics and time resolution performance of SiC detectors, a  simulation tool--RASER(RAdiation SEmiconductoR) is developed\cite{ref-RASER}.
Firstly, Based on an open source software, DEVSIM is used to simulate the electrical performance including I-V and C-V characteristics. Different from the commercial simulation software, DEVSIM has strong expandability and is easy to interact with Geant4 and other software for detector simulation. But the DEVSIM is in the primitive development, all the finite element equations should be written by users. As a TCAD device simulation package written in C++, with a Python front end, DEVSIM is capable of simulating 1D, 2D, 3D structures with models describing advanced physical effects and uses the control volume approach for assembling partial-differential equations (PDE’s) on the simulation mesh.\par
Secondly, RASER based on the ROOT, FEinCS and Geant4, has been used to simulate the time resolution of 4H-SiC PIN\cite{ref-sic_pin} and 3D 4H-SiC detector\cite{ref-3D}. In this work, RASER is also used to simulate the time resolution of 4H-SiC LGAD. When an incident-charged particle passes through the detector, it deposits a portion of its energy, producing an electron-hole pair. The electron-hole pairs generate in the detector move towards the poles in response to the electric field, creating an induced current at the electrodes. This current can be calculated using Shockley-Ramo's theorem:
\begin{equation}
I(t)=-q \overrightarrow{v}(\overrightarrow{r}(t)) \cdot \overrightarrow{E}_{w}(\overrightarrow{r}(t))
\end{equation}
where $\overrightarrow{r}$ is the path of electron or hole drift, $\overrightarrow{v}(\overrightarrow{r})$ is the drift velocity, $\overrightarrow{E}_{w}(\overrightarrow{r})$ is the weighting potential of $\overrightarrow{r}$. And the sum of induced current generated by the electrons and holes is the total current.

\subsection{The simulation of 4H-SiC PIN detector}
RASER based on the open source software DEVSIM can be used to simulate 4H-SiC detectors. In order to demonstrate the reliability of the RASER simulation tool, the 4H-SiC PIN detector\cite{ref-sic_pin} is selected as an example to compare the simulation results with the experimental results. Firstly, the size of the 4H-SiC PIN device under investigation is 5mm$\times$5mm. It has an active epitaxy layer with a thickness of 100$ ~\mu m$ and a doping concentration of 5.2×10$^{13}$cm$^{-3}$, and both of the top and bottom electrodes are ohmic contacts. The comparison of C-V performance between experimental result and the DEVSIM simulation result is shown in Figure 3(a). The good agreement between simulation result and experimental result indicates that the geometry and doping input added in the DEVSIM simulation are correct. Furthermore, for the IV simulation of the 4H-SiC PIN detector, considering the influence of tunnelling effect, the Hurkx Model has been added into the Generation \& Recombination (G\&R) constant model, and the simulated results are closer to the experimental results. Under the high electric field, the simulated results match well with the experimental results, but there are still significant differences under low electric field. This may be due to the complex defect features of SiC materials, which require further research. Based on the above research, the coefficients of the Hurkx model has been mathematically corrected, but the physical reasons are still unclear and require further study. The combination of IV simulation results and accurate CV simulation results indicates that RASER is capable of basic device simulation and has a certain reliability.

\begin{figure}[H]
\centering
\includegraphics[trim=0 40 0 20 ,scale=0.48]{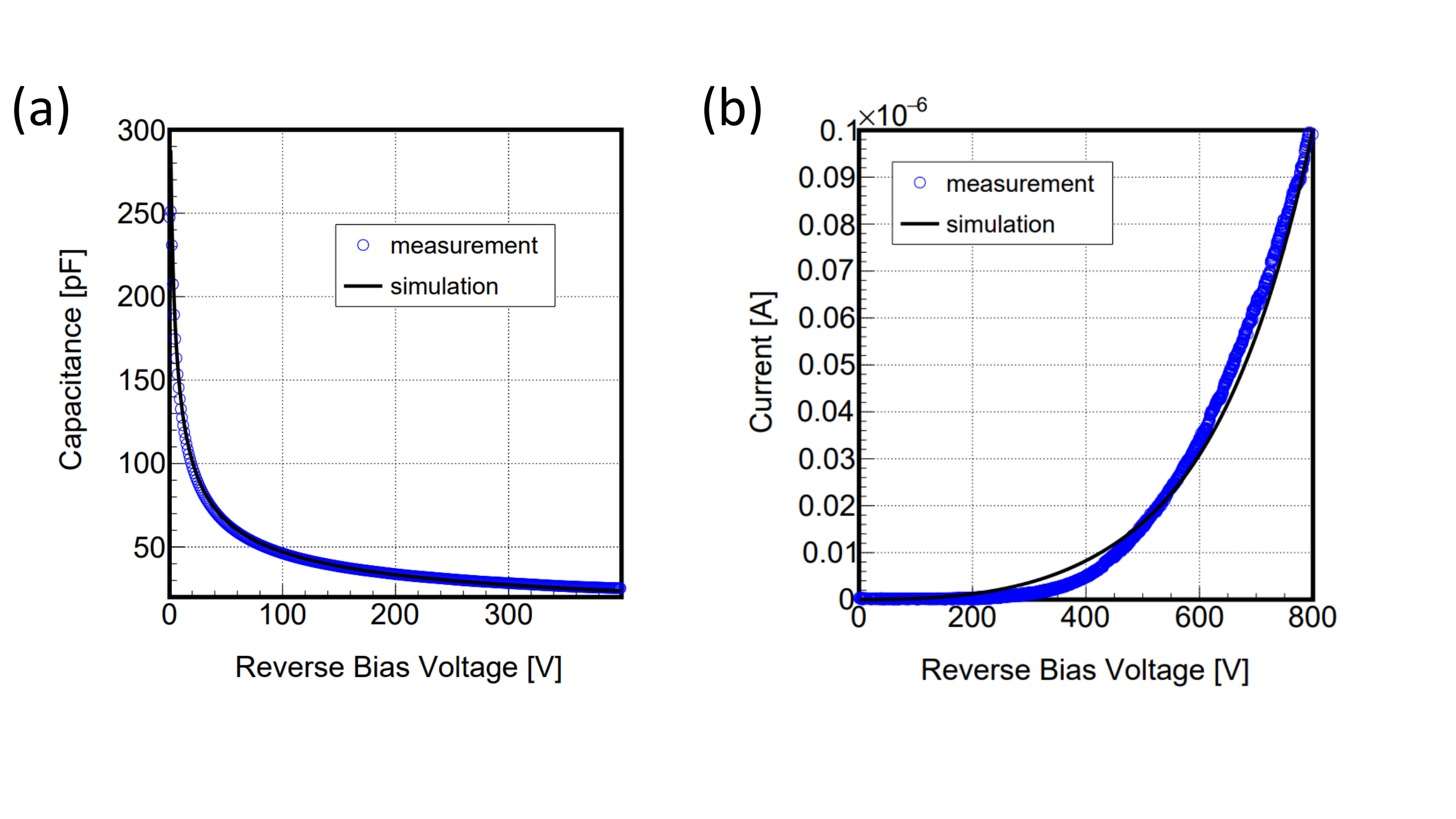}
\caption{Comparison between DEVSIM simulations and experimental results (a) C-V performance of DEVSIM simulation and experimental result (b) I-V performance of DEVSIM simulation and experimental result \label{figX}}
\end{figure}   
\unskip

In addition, the influence mechanism of some deep level defects on the leakage current was also studied. Two kinds of deep level defects, Z$_{1/2}$ and EH$_{6/7}$, in 4H-SiC materials were mainly investigated\cite{ref-deep level defects}. For Z$_{1/2}$, the defect concentration ranges from 1×10$^{13}$ cm$^{-3}$ to 1×10$^{15}$ cm$^{-3}$, and the capture cross section ranges from 1×10$^{12}$ cm$^{2}$ $\sim$ 1×10$^{16}$ cm$^{2}$. However, there is no significant effect on the leakage current from Z$_{1/2}$, which may be due to its relatively low energy level. On the other hand, for EH$_{6/7}$ with an energy level of 1.25 eV to 1.73 eV, the leakage current increases slightly with the increase of defect concentration shown as Figure 4(a). And the change of the capture cross section has little effect on the leakage current shown as Figure 4(b). From the simulation, the deep energy level defect has a slight effect on the leakage current of 4H-SiC PIN device, demonstrating it is not the dominant factor. Furthermore, it has been reported that macroscopic defects would greatly affect the carrier lifetime of the SiC devices, thus affecting the leakage current\cite{ref-macroscopic1,ref-macroscopic2}. Therefore, it is particularly important to avoid macroscopic defects in the preparation process of SiC detectors.

\begin{figure}[H]
\centering
\includegraphics[trim=0 40 0 20 ,scale=0.48]{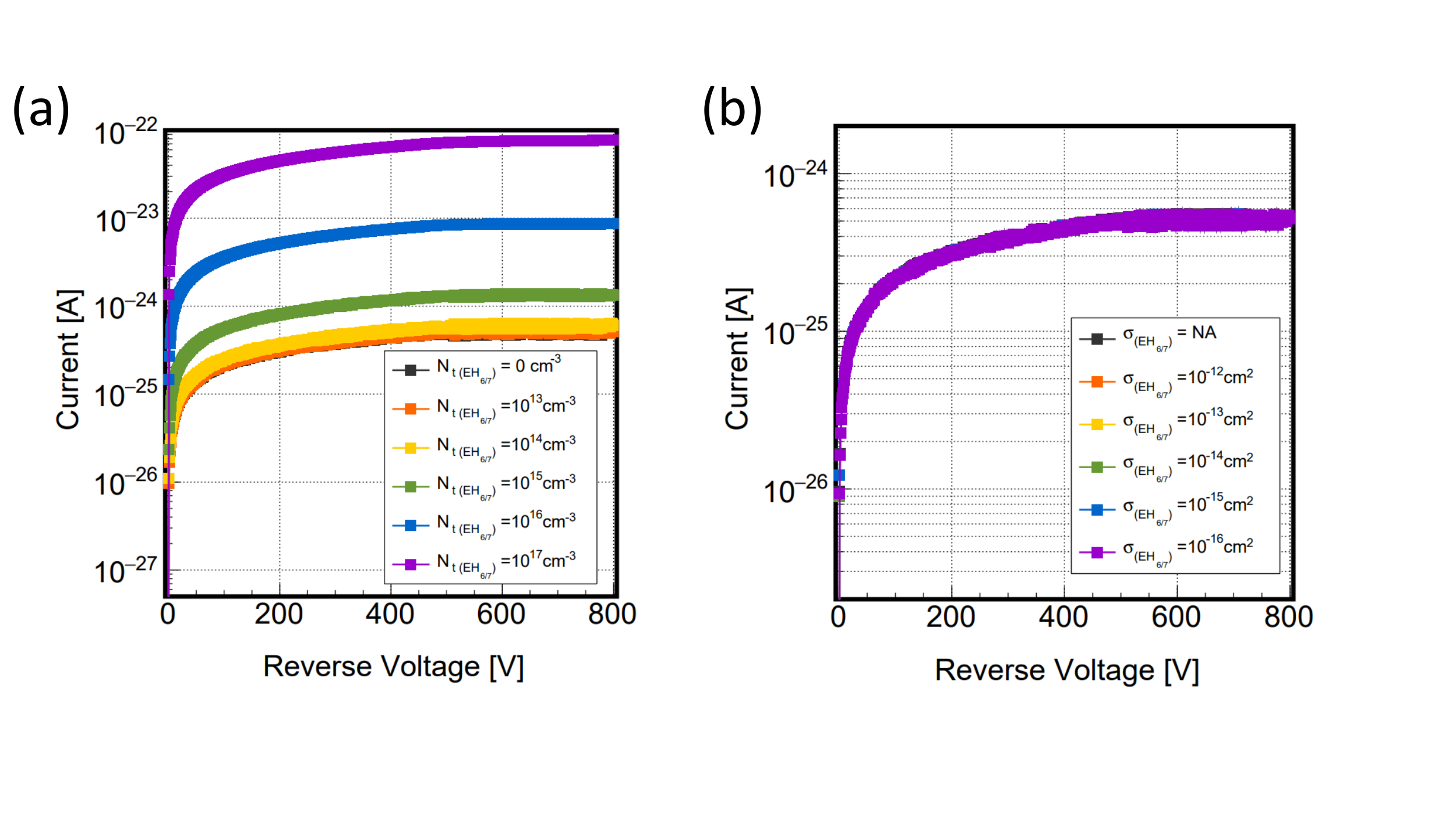}
\caption{(a) Effect of different $EH_{6/7}$ defect concentrations on leakage current. (b) Effect of capture cross-section on leakage current.}
\label{figX}
\end{figure}   
\unskip

\subsection{The simulation of 4H-SiC LGAD detector}
 
RASER was used to simulate the I-V and C-V curves of the 4H-SiC LGAD. As shown in Figure 5(a), the breakdown voltage of this device can be obtained from the I-V curve, which is approximately 3700 V of the 4H-SiC LGAD. The C-V curve, shown as the Figure 5(b), demonstrates the depletion voltage of gain layer V$_{GL}$ and the full depletion voltage V$_{FD}$. In detail, different from the PIN device, the C-V curve of the LGAD device shows a platform at around 130 V, which indicats that the gain layer of the device can be completely depleted at around 130 V. In addition, the full depletion voltage of the 4H-SiC LGAD obtained from Figure 5(b) is 400 V. Combined with the breakdown voltage, it can be considered that the working voltage range of the 4H-SiC LGAD is 400 V$\sim$3700 V. \par
Figure 5(a) and Figure 5(b) show the I-V and C-V curves of LGAD and PIN, respectively. These two devices are identical in structure and doping, except that the PIN has no gain layer. In terms of the IV curve, the breakdown voltage V$_{BD}$ of 4H-SiC LGAD is about 3700 V, it is smaller than that of PIN which does not breakdown until the reverse voltage is 4000 V. The reason for the phenomenon is that the high doping concentration of the gain layer in LGAD created an electric field peak, which makes easier for the carriers to accelerate and reach the breakdown energy. And at the same voltage, the leakage current of LGAD is larger than that of PIN due to the presence of the gain layer in LGAD. However, in terms of the CV curve, LGAD has one more inflection point than PIN. Different from the PIN devices, with the increase of the reverse voltage, the gain layer of the LGAD device is depleted before the bulk layer. \par
The IV and CV curves of PIN and LGAD we simulated show that, unlike the structure of PIN, LGAD has a gain layer and low gain multiplier effect. The structure of LGAD that we have designed is reasonable.

 \begin{figure}[H]
\centering
\includegraphics[trim=60 40 0 20 ,scale=0.38]{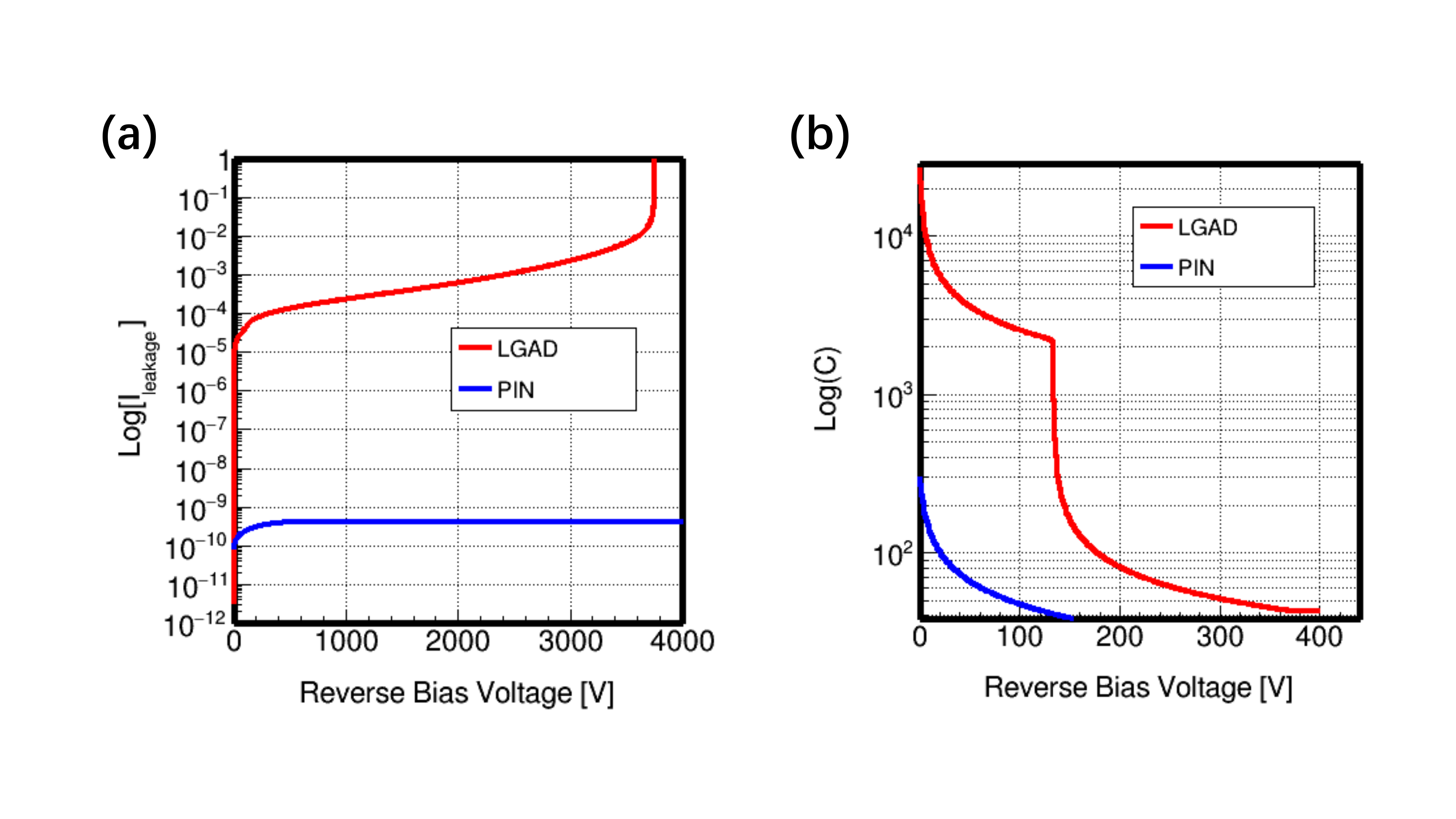}
\caption{The (a) current - voltage characteristics and the (b) capacitance - voltage characteristics of 4H-SiC LGAD and 4H-SiC PIN, used DEVSIM simualtion.\label{fig1}}
\end{figure}   
\unskip

\subsection{The time resolution simulation of 4H-SiC LGAD detector}
The time resolution of a detector is deﬁned by the accuracy of the measured arrival time of a detected particle. It can be calculated by the following formula \cite{ref-tr}:  

\begin{equation}
\sigma^2_t=\sigma^2_{TimeWalk}+\sigma^2_{Landau}+\sigma^2_{Distortion}+\sigma^2_{Jitter}+\sigma^2_{TDC}
\end{equation}
where the terms $\sigma_{TimeWalk}$ and $\sigma_{Landau}$ is from the random energy deposition of the detected particle, the physical process correspoding to which is simulated by Geant4. The term $\sigma_{TimeWalk}$, or the uncertainty of signal exceeding given threshold, can be eliminated by the constant fraction discrimination(CFD) method. 
The term $\sigma_{Distortion}$ is caused by the fluctuation of carrier velocity, which could be calculated from the Langevinian random motion of the carriers.
Noise in the circuit contributes to the term $\sigma_{Jitter}$. Since we set the electronics the same as \cite{ref-sic_pin} in the simulation, a Gaussian noise, parameters of which measured from \cite{ref-sic_pin}, is used.
The $\sigma_{TDC}$ is introduced by analog-to-digital converting, and is often small enough to ignore.\par
We simulated 50,000 events of beta particle detecting of the SiC LGAD detector under -800V bias voltage, and counts the time of arrival(ToA) of the particle to obtain the time resolution, which is (35.0 ± 0.2) ps, shown as the Figure 6(a). 
We also simulated beta particle detection under different detector working voltages. As figure 6(b) shows, the time resolution of the 4H-SiC detector decreases with the increasing of the reverse voltage. This is due to the fact that the drift velocity of the carriers increases as the voltage increases, resulting in the carriers being collected at a faster rate.

\begin{figure}[H]
\centering
\includegraphics[trim=0 40 0 20 ,scale=0.48]{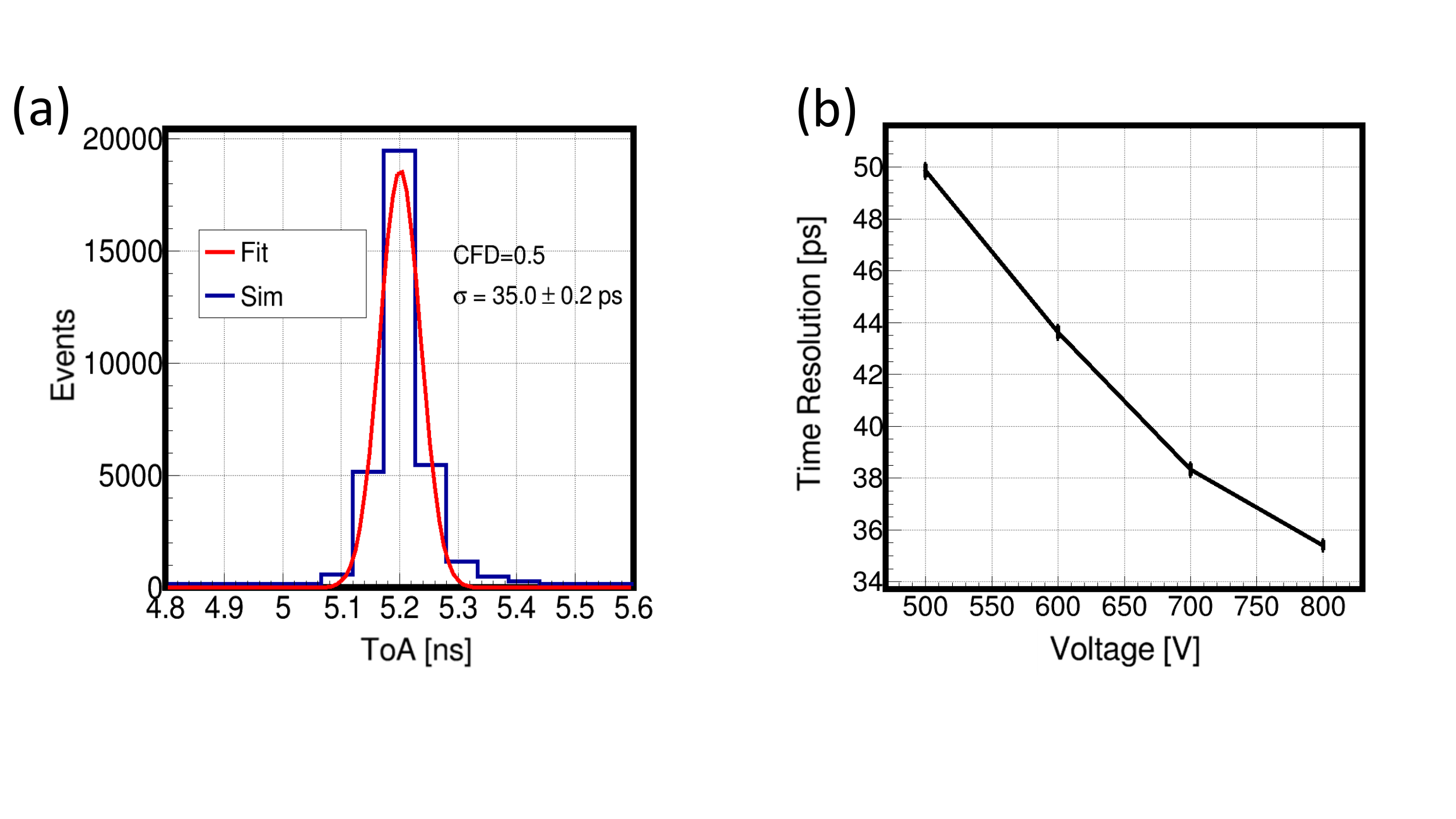}
\caption{The time resolution simulation results of the LGAD detector (a) when the reverse bias voltage is 800V and (b) the reverse bias voltage from 500V to 800V. \label{figX}}
\end{figure}   
\unskip

\section{Conclusion}\label{sec13}

In conclusion, the 4H-SiC LGAD has been designed including the epitaxial structure and technical processing and is currently in production. The electrical performance and time resolution of 4H-SiC LGAD has simulated by adjusting doping concentration and device structure using an open source software DEVSIM. Combined its I-V with C-V simulation results, the device has a full depletion voltage of 400 V and a breakdown voltage of 3700 V, which means that its operating range is 400 V$\sim$3700 V. Furthermore, the corresponding simulation exhibited that the device has a time resolution of (35.0 ± 0.2) ps at a reverse bias voltage of 800 V, exceeding that of the SiC PIN detector. This suggests that the 4H-SiC LGAD detector has the potential to be used in high irradiation environment at high temperature. And this work paves the way for the future development of 4H-SiC LGAD and its application in an extreme environment.

\backmatter

\bmhead{Acknowledgments}

This work is supported by the National Natural Science Foundation of China (No. 11961141014 and No. 12205321), China Postdoctoral Science Foundation (2022M710085), the State Key Laboratory of Particle Detection and Electronics (No. SKLPDE-ZZ-202218 and No. SKLPDE-KF-202313), Natural Science Foundation of Shandong Province Youth Fund(ZR202111120161) under CERN RD50 Collaboration framework.

\end{document}